\documentclass[a4paper,fleqn,usenatbib]{mnras}
\usepackage{natbib}
\usepackage{newtxtext,newtxmath}

\usepackage[T1]{fontenc}
\usepackage{ae,aecompl}

\usepackage{graphicx}	
\usepackage{amsmath}	
\usepackage{amssymb}	
\usepackage{multicol}        
\usepackage{mathtools}
\usepackage[usenames, dvipsnames]{color}

\DeclarePairedDelimiter\abs{\lvert}{\rvert}

\newif\ifdraft
\draftfalse

\title[Non-linear galaxy bias from phase differences]
{Obtaining non-linear galaxy bias constraints from galaxy-lensing phase differences}

\author[M. Manera]{
Marc Manera,$^{1,2}$\thanks{E-mail: m.manera@damtp.cam.ac.uk}
David Bacon$^{3}$
\\
$^{1}$Institut de F\'isica d'Altes Energies, The Barcelona Institute of Science and Technology, Campus UAB, 08193 Bellaterra (Barcelona), Spain\\ 
$^{2}$Kavli Institute for Cosmology, University of Cambridge, Madingley Road, Cambridge CB3 0HA, UK\\
$^{3}$Institute of Cosmology \& Gravitation, Dennis Sciama Building, University of Portsmouth, Portsmouth, PO1 3FX, UK\\
}

\date{Accepted XXX. Received YYY; in original form ZZZ}

\pubyear{2020}

\begin{document}
\label{firstpage}
\pagerange{\pageref{firstpage}--\pageref{lastpage}}
\maketitle

\begin{abstract}
We demonstrate the utility and constraining power of a new statistic for investigating galaxy bias: the galaxy-lensing
phase difference. The statistic consists in taking the differences 
of the phases of the harmonic wave-modes between the weak lensing convergence field and the galaxy
count field. We use dark matter simulations populated with galaxies up to redshift $z=1$ to 
test the performance of this estimator. We find that phase differences are sensitive to the absolute 
value of the second order bias ($c_2=b_2/b_1$), and demonstrate why this is the case. 
For a \~ 1500 sq. deg. galaxy survey we recover $c_2$ with an error of approximately $0.1$
for a wide range of $c_2$ values; current constraints from redshift surveys have errors of 0.1-0.6 depending on redshift.  
This new statistic is therefore expected to provide constraints for $c_2$ which are
complementary and competitive with constraining power by the conventional estimators 
from the power spectrum and bispectrum. For the Dark Energy Survey (DES), 
we predict leading measurements of second-order bias.  
\end{abstract}




\section{Introduction}

Galaxies are biased tracers of the matter field. In the standard cosmological paradigm, 
they form inside dark matter halos that have grown in the cosmic web from 
primordial seeds of inflation.
Galaxies, however, do not trace the matter density field perfectly, but do so in a non-linear, stochastic and 
environment-dependent way \citep[e.g.][]{2018PhR...733....1D}. The relation between the galaxy distribution and the underlying density distribution is governed by galaxy bias.

Galaxy bias can be measured from galaxy surveys by comparing the strength of the observed 
galaxy clustering (for instance, via the galaxy power spectrum or correlation function) to the 
strength of the matter clustering, either modelled by theory or obtained from 
weak lensing statistics.  The latter requires the measurement of galaxy ellipticities, which are used as estimates of the lensing shear. The cross-correlation between shear and galaxy counts (known as galaxy-galaxy lensing) can also inform about bias \citep[see e.g.][]{2018PhRvD..98d2005P}.

Galaxy bias may be separated into a linear (i.e. multiplicative) constant, which dominates
at large scales, and non-linear contributions that dominate at smaller scales. While the
linear galaxy bias is relatively easy to measure, involving only the 2-point statistics, measuring the non-linear bias takes considerably more effort
as it usually requires the use of higher order clustering estimators such as the 3-point
correlation function or the bispectrum. 
 
Measurements of non-linear galaxy bias from galaxy clustering have been performed 
from cosmological surveys, including APM \citep{1994ApJ...437L..13G}, 2dF \citep{2002MNRAS.335..432V}, and BOSS \citep{2015MNRAS.451..539G}.
Current results gives constraints on the non-linear bias $b_2\simeq 0.6$ with errors of between 0.07 and 0.26, depending
on the redshift and galaxy sample \citep{GilMarin2017}.
 
In this paper we demonstrate the utility and constraining power of a new statistic 
to measure the non-linear galaxy bias. The statistic consists in taking the differences 
of the phases of the harmonic wave-modes between the weak lensing-derived convergence field 
and the galaxy count field.  This approach takes advantage of the wealth of non-Gaussian information contained 
in the phases without having to measure the three-point function or bispectrum. 

Notice also that the galaxy bias in our method is directly measured from the weak lensing and galaxy fields,
without the need to assume a theoretical dark matter clustering strength.

The phases of the dark matter field and their evolution have previously been studied by
\cite{ChiangNaselsky2004},
\cite{ChiangColes2000},
\cite{ChiangColes2002},
\cite{WattsColes2003},
and 
\cite{WolstenhulmeBoyin2015}. 
In \cite{SzepietowskiBacon2014} the difference between the phases of the matter and galaxy fields
was used as a weak prior to reconstruct convergence maps. 

The effectiveness in constraining cosmology with a 
three point phase correlation function, the line correlation function (LCF), 
has been investigated using Fisher forecasts by \cite{ByunLCF} and \cite{Eggemeier2017} finding an improvement 
of cosmological parameter constraints by about 20\%.  
When using the LCF in combination with the power spectrum 
the marginalized errors on $b_2$ improve by 13-33\%, depending on the scales. 
Using simulations, \cite{Kamran2018} also found that the LCF 
improves parameter constraints for non-LCDM cosmologies. 

In this paper, the difference between the phases of galaxies and dark matter fields 
is explored in order to provide a new method to constrain galaxy bias. 
We simulate the weak lensing and galaxy clustering of cosmological galaxy samples to redshift $z=1$, for a region
comparable to that observed by the Dark Energy Survey. We show that non-linear galaxy bias 
$c_1$ can be measured to an accuracy of 0.1 using phase differences, for a wide range of $c_2$ values.    
This paper is therefore a proof of concept that the phase difference statistic has good constraining power.

\section{Theory}
	\label{theory}

\subsection{Galaxy Bias} 

Galaxy bias can be modelled at several levels of complexity. 
The simplest model considers 
galaxies as perfect but discrete tracers of the dark matter field. 
That is, the expected number of galaxies in a given cell volume, $V$,
is proportional to the total amount of matter, $\rho V$, in that volume, where $\rho$ is the local matter density in this volume. This 
may be expressed as
\begin{equation}
E[n_g](x) = f\, V \rho (x) , 
\end{equation}
where $E$ indicates expectation, $f$ is a proportionality constant, 
and $x$ stands for the centre position of the chosen volume.  

A second step of complexity consists in modelling the galaxy bias
as a non-linear but local function. For instance,
\cite{delaTorre2013} chose a power law relating bias and mass for their simulated galaxy catalogue:
\begin{equation}
E[n_g](x) = A\, \rho(x)^{b_p}.  
\end{equation}
A more common approach is to work within the Perturbation Theory (PT) 
framework. In this case the natural quantities are the density fluctuations of matter,
$\delta(x) = \rho(x)/\bar{\rho} - 1$, and the density fluctuations of galaxies, 
$\delta_g(x) = E[n_g](x)/\bar{n}_g - 1$, computed using the mean
density of matter $\bar{\rho}$ and of galaxies $\bar{n}_g$
in cells of volume $V$. Limiting ourselves to local terms
up to second order, we can write \citep[e.g.][]{BelHoffmannGazta15}
\begin{equation}
\delta_g = b_1 \left\{ \delta_m + \frac{c_2}{2} ( \delta_m^2(x) - \sigma_m^2 ) + \mathcal{O}(\delta^3) \right\}, 
\label{biasPT}
\end{equation} 
where $b_1$ is the linear bias, $c_2$ is a measure of the relative second order bias, and
$\sigma_m^2$ is the variance of $\delta_m$, which has been introduced to keep $<\delta_g> = 0$. 
Notice that the density fluctuations are evaluated in cells of a given volume; consequently
the bias parameters will depend on the smoothing scale set by that volume. 
Further extensions are possible by including non-local bias terms 
\citep{Chan12,Baldauf12,BelHoffmannGazta15} which we will neglect in this paper.  

\subsection{Galaxy Noise}
 
The actual number of galaxies observed at a given position in our
Universe, $n_g$, will not be exactly equal to the expected number of galaxies, $E[n_g]$, 
but will be a random integer variable drawn from a distribution with the
expected number as its mean, i.e, 
\begin{equation}
n_g(x) = E[n_g](x) + \epsilon_g(x), 
\end{equation}
where $\epsilon_g(x)$ is the noise contribution, and $E(\epsilon_g)=0$.
We use a Poisson distribution for the noise. 

\subsection{Weak Lensing}

Weak lensing refers to slight distortions in galaxy shapes, sizes and magnitudes
that occur due to  light having a bent trajectory due to perturbations in the
gravitational potential. The magnitude of the dilation of an image in the presence of lensing is given  
by the convergence, $\kappa$, which is a weighted projection of the matter fluctuations
on the line of sight. For galaxies at a given redshift, $z$, the convergence
is 
\begin{equation}
\label{km}
\kappa(\theta,r'(z))=\int_0^{r'(z)} dr q(r,r') \delta(\theta,r),
\end{equation}
where $\theta$ is the line of sight and $r(z)$ is the radial comoving distance.
Here the weak lensing projection kernel $q(r,r')$ is given by \citep{2001PhR...340..291B}
\begin{equation}
q(r,r') = \frac{3 H_0^2 \Omega_m}{2 c^2} \frac{r(r'-r)}{r' a(r)},
\label{eq:q}
\end{equation}
where $c$ is the velocity of light, $a$ the scale factor, 
$\Omega_m$ the density of matter in the Universe in terms of the critical density, 
and $H_0$ is the Hubble constant. 

Since we are interested in measuring galaxy bias from the phase information of the
weak lensing and galaxy fields, it is convenient to define a quantity that
resembles the weak lensing convergence, but contains information about the field galaxy count. That is,
we consider two samples of galaxies, the source galaxies that are at redshift $z$,
from which weak lensing measurements are taken, and the field galaxies, covering the volume
from redshift zero to $z$, which trace the matter, and for which we want to
obtain the galaxy bias. Consequently, for convenience, 
we construct the `galaxy convergence', $\kappa_g$ as follows: 
\begin{equation}
\kappa_g(\theta)=\int_0^{r'(z)} dr q(r,r') \delta_g(\theta,r),
\label{kg}
\end{equation}
where $q$ is given by equation (\ref{eq:q}). If the galaxy field were to trace matter perfectly, without noise and with the same constant of proportionality at all redshifts, then
$\kappa_g$ will be proportional to $\kappa$, and the Fourier or harmonic phases of the two fields would be
the same. In reality however, these conditions are not true and so differences in the Fourier or harmonic phases exist. They bring, 
therefore, an opportunity to measure non-linear galaxy bias.  

\subsection{Shape noise}

Weak lensing measurements are statistical in nature. Each galaxy has its ellipticity measured, typically using an ellipticity fitting method \citep[e.g.][]{2018MNRAS.481.1149Z}. These ellipticities are then used as estimators of the weak gravitational shear. But since each galaxy has its
own intrinsic shape, weak lensing convergence maps from real ellipticity measurements 
have some shape noise, which we model in pixels as 
\begin{equation}
\epsilon_\kappa(\theta,z) = \frac{\sigma_\kappa (\theta,z)}{\sqrt{n_s(\theta,z)}}, 
\end{equation}
where $\sigma_\kappa$ is the expected intrinsic shape contribution of a typical source galaxy
at redshift $z$, and $n_s$ is the number of source galaxies contributing to the convergence 
measurement of a given pixel.  
For our modelling we choose $\sigma_\kappa = 0.3$ and $n_s = 10$ at $z = 1$, 
which are values comparable to those of galaxy samples from the current cosmological lensing surveys \citep[][Table 5]{2018MNRAS.481.1149Z}.

\subsection{Phase differences}

In order to compute the phase differences between the galaxy and weak lensing convergence 
fields, we need to express these fields as a set of coefficients in a basis of orthogonal 
(or orthonormal) functions, which we take to be the Fourier space in the plane or the 
harmonic space on the sphere. In this paper we choose to work in harmonic space, since
our galaxy and convergence fields are on the celestial sphere and we want to model
galaxy surveys that cover a considerable fraction of the sky. For smaller areas, a
plane parallel approximation would be adequate, and an equivalent formulation in Fourier
space is possible. 

A function $f$ on the sphere can be expanded in spherical harmonics, $Y_{lm}$, as  
\begin{equation}
f(\theta,\varphi) = \sum_{l=0}^{\infty} \sum_{-l<m<l} a_{lm} Y_{lm}(\theta,\varphi),
\end{equation}
where $(\theta, \varphi)$ is the 2D position in spherical coordinates, and $a_{lm}$
are the harmonic coefficients, which are complex numbers that may be expressed
as a modulus and a phase,
\begin{equation}
a_{lm} = \abs{a_{lm}} \exp(i\phi_{lm}).    
\end{equation}
Taking the difference of the phases between the galaxy and weak lensing convergence
fields, for a particular $(l,m)$, gives our quantity of interest,
\begin{equation}
\Delta\phi_{lm} = \phi_{lm}^\textrm{galaxies} - \phi_{lm}^\textrm{lensing}.
\end{equation}
Note that for spherical harmonics, $Y_{lm}$ with $m=0$ do not 
have an azimuthal variation and hence $a_{l0}$ are real; 
consequently there is no phase difference between fields for these modes. 
We therefore do not include $m=0$ modes in our analysis.
In addition, since both the galaxy and weak lensing 
convergence fields are real, the harmonic
coefficients acquire a symmetry when changing the sign of $l$, $a_{lm} = a^{*}_{-lm}$, 
or equivalently,
$\Delta\phi_{l,m} = \Delta\phi_{-lm}$, thus we only use $l>0$ in our analysis. 
Finally, we choose a range of $l$, $20 < l < 200$, that covers the angular scales
from the size of our smoothing angular scale at the lowest redshift to about 
half the scale of the simulated survey.

\subsection{Sensitivity to the bias parameters}
\label{subsection:sensitivityc2}
Phase differences are sensitive to non-linear galaxy bias parameters but 
insensitive to linear galaxy bias. 

From Eqs. (\ref{km}) and (\ref{kg}) 
it is clearly seen that a linear galaxy bias 
would result in $\kappa$ and $\kappa_g$
being proportional to each other, modulo contributions from noise,
and therefore would have indistinguishable phase information. 
Although the value of the linear bias cannot be recovered from phase
information, it would still be recovered from 
galaxy clustering, where the modulus of the wave modes is measured. For example, comparing the angular power of spectrum
of $\kappa_g$ and $\kappa$ at lower wavenumbers would give the linear bias.

Regarding the sensitivity to the non-linear galaxy bias, we find 
the phase difference to depend on the modulus of the hierarchical bias
parameter $c_2$, where $c_2 = b_2/b_1$. This can be seen by 
explicitly writing the phase difference between $\kappa_g$ and
$\kappa$. Let us start with the harmonic coefficients of $\kappa$,
which are defined as:
\begin{equation}
a_{lm} = \int Y^{*}_{lm} (\theta) \kappa(\theta) d\theta
\end{equation}
where $\theta$ is the position in the sky. 

Using the properties of the spherical harmonics, we find the harmonic coefficients
of $\kappa_g$ for $l\ne 0$ to be as follows:
\begin{equation}
a^{g}_{lm} = b_1 a_{lm} + \frac{b_2}{2} \sum C_{l,l',l'',m,m',m''}  a_{l'm'} a_{l''m''}
\end{equation}
where $C$ are the Clebsch-Gordan coefficients for SO(3), 
which are functions of 3-j symbols, and the sum is over the values of $l',l'', m',m''$ allowed by the addition 
rule of the group. This sum is the equivalent of the convolution of Fourier fields in the plane. 
The exact values do not matter for our purposes as we only want the dependence 
of the phases on $b_1$ and $b_2$. 

Now, each harmonic coefficient has a real and imaginary part. Using the simplified notation 
in which $\psi'_1 = \textrm{Re}(a_{l'm'})$ and $\psi'_2 = \textrm{Im}(a_{l'm'})$ we
can express the phase of the galaxy convergence field as follows: 

\begin{equation}
\phi^{gal}_{lm} = \textrm{arctan}\left( \frac
{b_1 \psi_2 + \frac{b_2}{2} \sum C (\psi'_1\psi''_2 + \psi'_2 \psi''_1 )} 
{b_1 \psi_1 + \frac{b_2}{2} \sum C (\psi'_1\psi''_1 - \psi'_2 \psi''_2)} \right), 
\label{eq:symmetry}
\end{equation}
where we have dropped the indices on the sum and on the $C$ coefficients. Finally, knowing that
$\arctan(x)-\arctan(y)=\arctan((x-y)/(1+xy))$, we can compute the
difference between $\phi^{gal}_{lm}$ and $\phi^{lens}_{l,m}=\arctan({\psi_2/\psi_1})$, obtaining our 
quantity of interest

\begin{equation}
\Delta\phi_{lm}=\phi^{gal}_{l,m} - \phi^{lens}_{l,m}= 
   \arctan{\left( \frac{\frac{c_2}{2} S_x}
{\psi_1^2  + \psi_2^2 + 
               \frac{c_2}{2} S_y}\right)}
\label{dphic2}
\end{equation}
where 
\begin{equation}
S_x = \sum C ( \psi_1 \psi'_1\psi''_2 + \psi_1 \psi'_2 \psi''_1  
                      - \psi_2 \psi'_1\psi''_1 + \psi_2 \psi'_2 \psi''_2) 
\end{equation}
\begin{equation}
S_y = \sum C ( \psi_1 \psi'_1\psi''_1 - \psi_1 \psi'_2 \psi''_2 
                                    + \psi_2 \psi'_1\psi''_2 + \psi_2 \psi'_2 \psi''_1)    
\end{equation}

It is clear that the phase difference depends only on $c_2$. 
In addition, at leading order in $\psi$, the symmetries in Equation (\ref{dphic2}) imply that
we cannot distinguish between a positive and a negative $c_2$. Two different contributions
result in this effect. First, if the linear density field is Gaussian, there is a symmetry
between positive and negative densities. Flipping the sign of the density field will result
in a matter realization with the same cosmological parameters but in which all $\psi$'s will
have opposite signs. The phase differences will be the same as having changed the sign of
$c_2$. Secondly, even if the matter field is non-Gaussian, we can explore the symmetry of
reversing the azimuthal angle $\varphi$. Two maps related by this symmetry will have the
same underlying cosmology; at the same time,  
the imaginary parts of the spherical harmonics coefficients 
will have opposite signs: $\psi_2 \rightarrow -\psi_2$. At leading order in $\psi$, this 
change of sign in the imaginary parts can be compensated by a change in sign in $c_2$. 
As a consequence of this symmetry we are only sensitive to the modulus of $c_2$. 
In Appendix \ref{APc2} we demonstrate the $c_2$ symmetry in more detail and show relevant likelihood plots. 
In practice, being sensitive only to the modulus of $c_2$ means that we need prior knowledge to
select between positive and negative values. In general, this is no problem as we have
theoretical derivations of the galaxy bias from the peak-background split as well as fits to simulations.
Typically, we expect $c_2 < 0$ for $b_1 \lesssim 2$ and $c_2 > 0$ otherwise \citep{LaBaldauf2016,Hoffmann2017}.

\section{Simulated Data Samples}
\label{sims}

We simulated galaxy catalogues for our analysis using
the L-PICOLA code \citep{lpicola15}.  L-PICOLA rapidly produces
dark matter fields by splitting the gravitational 
equations into small and large scales and applying a 
combination of Particle-Mesh evolution and 2nd Order
Lagrangian Perturbation Theory. 
L-PICOLA is an efficient parallelization of the COLA algorithm 
\citep{cola13}. In addition to dark matter comoving outputs, L-PICOLA also includes 
dark matter outputs on the light-cone, and an option for primordial non-Gaussianities
inherited from the PTHalos code of \cite{pthalos13,pthalos15}. 
Because L-PICOLA is orders of magnitude faster than N-body codes,
it enables us to generate the required volume for our statistical analysis. 

We have created two fast
dark matter simulations in the lightcone up to redshift $z=1$,
with a base cubic box of size $\textrm{L}=4800\, \textrm{Mpc h}^{-1}$. 
Each box has $\sim2.6\times 10^{11}$ particles ($\textrm{N}_p=6400^3$)
and covers the full sky with an observer at the centre. 
The particle mass is $M_p = 1.09 \times 10^{10} M_{\odot}/h$. 
The dark matter is stored in $224$ concentric  
Healpix maps \citep{2005ApJ...622..759G}
of $N_{side}=4096$. Each Healpix pixel area is then $0.74$ sq. arcmin.

We divided the full sky into 28 regions of equal area, each of
approximately $1473$ sq. deg. This area is similar to the first
year of observations of the Dark Energy Survey \citep{2018PhRvD..98d3526A}.
These areas are constructed as follows. First, we cut 
the celestial sphere through the equator and at $\pm 40$ deg in declination. 
This gives four circular regions. The two regions closest to the equator
we divide into nine areas each by cutting at intervals of 40 deg in right ascension.
The two regions touching the poles we divide into five areas each by cutting
at intervals of 72 deg. in right ascencion. Since we have two full sky lightcone
simulations this yields a total of 56 areas, each with an area of $1473$ sq. deg.\footnote{The difference between the two types of areas is only at the level of $\sim 0.03$ per cent.}
In this paper we use 48 of these regions, excluding 8 for technical reasons.\footnote{Results do not significantly change when including those regions but they were 
excluded as the dark matter distribution in a very small area in those regions was not stored.}

Using the relations in Section \ref{theory}, 
from the overdensity values in the Healpix matter maps
we have constructed maps for the weak lensing convergence, $\kappa$,
and the galaxy convergence $\kappa_g$, which are the basis of our analysis. 
Galaxies are assigned throughout the line of sight to 1.5\% of our 
dark matter particles, resulting in a number density of $\sim$10 per sq. arcmin.
Galaxies below $z < 0.05$ have not been included in our analysis because
of their large discretization effects, as explained in Appendix \ref{kg-appendix}.

For the contribution of the non-linear bias, $b_2$ to the galaxy overdensity we
have smoothed the dark matter Healpix maps with a Gaussian kernel 
equivalent to $2.5$ Mpc/h full width half maximum, to reduce the noise
in the dark matter field. 
Our choice of the maximum wavenumber $\ell$ takes into account this scale.  

The simulation cosmological parameters are $\Omega_m = 0.25$,
$\Omega_b = 0.044$, $\Omega_\Lambda=0.75$, $h=0.7$, $\sigma_8=0.8$
and $n_s = 0.95$.

\section{Statistics}
\label{statistics}

In this section we explain how we evaluate the probabilities
that we use for the likelihood plots, where
we show the constraining power of the phase differences.

Consider an observation of an area of the sky from a survey like DES,
from which we have the phase differences of a collection of wave modes.
We want to evaluate the average constraining power of such an observation 
for the galaxy bias parameters. For this, we need to estimate   
the probability that a particular observation is a realisation of
a given bias model with some values for its parameters.

A standard approach for this would be to compute histograms of the phase differences
in a set of wave mode bins, estimate their errors and covariances from simulations, 
and then fit the parameters of a given model to the data histograms, 
perhaps by minimizing the $\chi^2$. 

In our case, however, we do not have sufficient simulated realisations 
or volumes of our survey area to compute reliable covariance matrices 
for the set of phase difference histograms. To address this, one could
seek to evaluate the covariance matrix from a single survey volume 
using resampling techniques such as jackknife subsampling. Jackknife
and resampling techniques, however, are known to fail to estimate 
the variance of sample quartiles \citep{jackknifeFAILS,bootstrapbook}.
We choose therefore an alternative approach that consists in compressing
the relevant information into a particular statistic. 


For each wave-mode bin, we compute the Kolmogorov-Smirnov distance 
between the observed and the modelled phase difference histograms. 
The Kolmogorov-Smirnov distance, $d$, 
between two sets of numbers is the maximum difference between the normalized 
cumulative distribution of the sets. 
This is a useful statistic since,
for the case of a single variable, it can be mapped to the probability that
the two sets of numbers are drawn from the same distribution. 

For our analysis, we choose as our statistic a particular combination
of Kolmogorov-Smirnov distances between the set of phase difference values
of the observed data and the equivalent set from the simulated data with the
bias parameters fixed. The details are explained below. For the case
of analysing only one bin of wave modes, our chosen statistic can be simply understood as: the mean of the Kolmogorov-Smirnov distances between (i) a series of 
phase difference realisations with the parameter values chosen as our `observation' 
and (ii) a series of phase difference realisations with the bias parameters 
which we are comparing to our observation.  

Now our chosen statistic, $\mathcal{Q}$, is a measure of the distance
between the observations and the model. However, this distance by itself
doesn't give us a likelihood of our data given the model parameters; 
for this we need the equivalent of the covariance matrix, i.e, the intrinsic
dispersion of our statistic due to cosmic variance and stochastic effects.
We compute the likelihood by calculating how rare the value of our 
statistic is compared with the dispersion in the possible observations generated by a particular fiducial parameter set.

In this paper we are not yet using  observational data from a survey;
instead, as proxy for our observed universe,
we use the simulations set at given fiducial bias parameters.
In order to compute the intrinsic dispersion of our statistic, we then
use the 48 realisations of our survey area, fixed at the fiducial bias
values that we chose as corresponding to the observations. With this
we have all the elements required for the statistical analysis. 

\subsection{Statistical method}

Here we describe in detail our statistical approach.

\begin{enumerate}

\item We choose a bias model and, within this model, a set of variable parameters $\vec{\beta}$.
Our results are presented for the perturbation theory model of Eq \ref{biasPT}.
In this case, $\beta = (b_1,c_2)$ are our variable parameters. 

\item We choose a set of particular values of the variable parameters as our fiducial values.
We label them $\beta_0$. They represent the bias values of our `observable' universe. 
For instance, the top right panel of Figure \ref{likeliAA} 
shows a likelihood contour plot with fiducial values $b_1=1.35$ and $c_2=0.3$ as proxy for our observation.

\item We take our $n=48$ dark matter simulations of the survey volume (see section \ref{sims}) and populate each with galaxies according to  the fiducial bias parameter values chosen for our universe. 
Each realisation may be taken as a possible proxy for an observation from the survey.
\item For each of these 48 realisations we obtain the phase differences between the weak lensing
and galaxy convergence fields. The differences between realisations are due to cosmic variance
and stochastic noise. For each realisation we compute the mean of the Kolmogorov-Smirnov 
distances between that realisation and all the others, doing this for each bin of wave modes. 
We have nine bins of wave modes labelled by $\ell bin$. Consequently we have nine distributions of means of Kolmogorov
distances, \{$\mathcal{D}^{\ell bin}_i(\beta_0)\}$, where 
$\mathcal{D}^{\ell bin}_i (\beta_0)=\frac{1}{n-1}\sum_{i\neq j} d^{\ell bin}_{ij}$
and $d^{\ell bin}_{ij}$ is the Kolmogorov-Smirnov distance between 
the realisations $i$ and $j$ in the wave mode bin $\ell bin$.
The full distribution of realisations gives us a means of calculating 
a quantity equivalent to a covariance matrix, i.e, a way to obtain a probability of how far a 
model with a particular set of parameter values is from a universe with the fiducial observational
parameters.

\item We choose a flat prior for our variable model parameters, which we cover with a regular
grid of points. In our plots, our variable model parameters, $b_1$ and $c_2$, are set 
in a grid with a step size of $0.0125$, thus covering our parameter space with thousands of points.

\item At each of our points $\beta$ in the parameter space, we run 48 realisations of our model 
with the parameter values given by the point, each with the volume of the survey. 
These realisations have the same underlying dark matter as the fiducial 48 realisations
that we use as proxy for observations, but in this case the galaxies are populated with 
different bias parameter values (those according to each grid point) and with different noise. 
For each of the 48 realisations we compute the phase differences between the galaxy and the weak lensing
convergence.

\item  Now, at each grid point, $\beta$, we want to estimate the distance
between our model with its given parameter values, and our proxy for the observed universe 
with its fiducial parameter values. This distance, $\mathcal{Q}(\beta)$, is our chosen statistic.

\item  To estimate $\mathcal{Q}(\beta)$, let us start by focusing momentarily on only 
one of the proxy realisations of the observed universe.
For this realisation $k$, and for each $\ell bin$, we compute Kolmogorov-Smirnov
distance $d_{k,m}$ between this realisation with fiducial parameter values $\beta_0$ and each of
the 48 realisations (indexed by $m$) of our survey with our model parameter values $\beta$. Taking the 
average of the Kolmogorov-Smirnov values we have a measure of the distance between
the proxy realisation for an observation and a model with parameters $\beta$.  
Mathematically we can write the mean as follows:
$\mathcal{D}_k^{\ell bin}(\beta) = \frac{1}{n} \sum_{k} d_{k,m}^{\ell bin}$.

\item Now we have $\mathcal{D}^{\ell bin}_k(\beta)$, the mean distance between a particular fiducial observation $k$ and all the model realizations. We also have $\{\mathcal{D}^{\ell bin}_i (\beta_0)\}$, the set of mean distances
between the various observation realizations, telling us about the cosmic and noise variances inherent in our measurements. So we can assign a probability $P_k^{\ell bin}(\beta)$ that a distance as large as $\mathcal{D}^{\ell bin}_k(\beta)$ would be obtained by chance 
given the statistical variation in distances expected in our observations. We can estimate this probability as the fraction
of fiducial realizations with $\mathcal{D}^{\ell bin}_i (\beta_0) > {\mathcal{D}}^{\ell bin}_k(\beta)$. 

\item This probability is well behaved. If $\beta = \beta_0$, i.e, if the model we are
testing is the same as the observation, then the set of the survey realisations of the model
yields a Gaussian distribution. If the model is far away from the observations, then 
$\mathcal{D}^{\ell bin}_i (\beta_0) < \bar{\mathcal{D}}^{\ell bin}_k$ for almost all or 
all $i$, and the probability goes to zero. 

\item Up to this point our statistics come from a set 
of average internal distances within a set of possible fiducial observations $\{\mathcal{D}_i\}$
and the average distance between model realizations and one single observation $\mathcal{D}_k$.
Because this single observation might not be the most representative of $\beta_0$, using the distance to one single realisation of the observations
introduces cosmic variance. As a result, the likelihood values of our plots constraining
the parameter values would not be centred at $\beta_0$. 

\item A better estimation of the statistical
distance between a universe with the fiducial parameters $\beta_0$ and a model $\beta$
is to take the mean of the distances to each of the fiducial realisations of the survey.
Consequently, we set our distance measurement to be 
$\bar{\mathcal{D}}^{\ell bin} (\beta) = \frac{1}{n} \sum_k \mathcal{D}_k^{\ell bin} (\beta)
= \frac{1}{n^2} \sum_{k,m} d_{k,m}^{\ell bin}$. Therefore, we estimate the probability $P^{\ell bin}(\beta)$ 
that the value of the parameters of our model $\beta$ are consistent with our proxy universe as the fraction of distances 
$\{\mathcal{D}_i^{\ell bin}(\beta_0)\}$ that have values larger $\bar{\mathcal{D}}^{\ell bin}(\beta)$. 

\item We are ready now to declare our statistic. 
When we are interested only in one $\ell bin$, our statistic is ${\mathcal Q}(\beta)=P^{\ell bin}(\beta)$ 
and our likelihood at each point is proportional to this value.

\item When we are interested in a combination of wave mode bins, our statistic
is ${\mathcal Q}(\beta) = \prod_{\ell bins} P^{\ell bin}(\beta)$. This statistic is proportional
to the likelihood at each point in the parameter space, if there
are no correlations between the wave mode bins. 
However, as correlations exist we need to compare ${\mathcal Q}(\beta)$ 
(which can be considered to be the distance between the model with parameters $\beta$ and the observation) 
against a set of ${\mathcal Q}_i(\beta_0)$ values, which are the distances between 
the various realisations obtained from the parameters underlying our observation. 
In this case, ${\mathcal Q}_i(\beta_0) = \prod_{\ell bins} P_i^{\ell bin}(\beta_0)$
where $i$ stands for the set of observation realisations of our fiducial values.
These values take into account the correlation between $\ell bins$.

\item Finally, we can estimate the probability that our model parameters are consistent with our fiducial case, by 
comparing ${\mathcal Q}(\beta)$ to the distribution of $\{{\mathcal Q}_i(\beta_0)\}$. 
The likelihood of our data given the model is then proportional to the fraction of 
${\mathcal Q}_i(\beta_0) < {\mathcal Q}(\beta)$. 
In addition, given that our priors are flat, the posterior probability
is also proportional to the same quantity. 

\end{enumerate}

\section{results}
\label{sec:results}

\begin{figure}
 \includegraphics[width=250px]{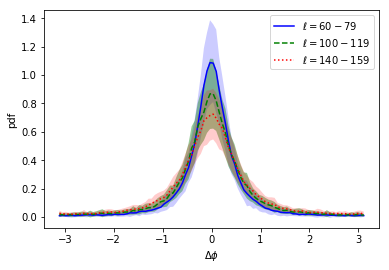}
 \includegraphics[width=250px]{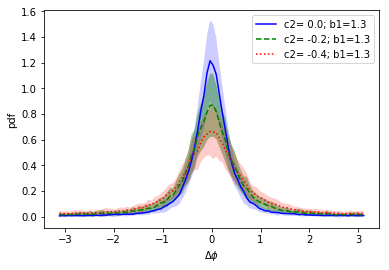}
 \includegraphics[width=250px]{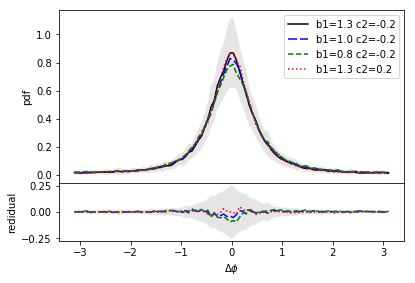}
 \caption{
   Top panel: phase difference histograms for different values of the 
harmonic coefficients, with bias parameters fixed 
($b_1 = 1.3$ and $c_2 = -0.2$). Middle panel: 
phase-difference histogram for different values of the nonlinear
bias $c_2$, with the linear bias, and
the harmonic bins fixed ($b_1=1.3$, $l={100-119}$). 
Bottom panel: phase difference histograms showing that only the absolute value
of the non-linear parameter matters in practice, 
regardless of the linear bias value. Here we kept $l={100-119}$.
In all panels, the coloured bands show the RMS errors from the set
of survey simulations.}
 \label{histograms}
\end{figure}

In this section we present results for a DES like survey with 
$1,473$ sq. deg. with galaxies up to redshift $z=1$. These 
results are the proof of concept that phase differences between
the galaxy convergence field, $\kappa_g$, and 
the weak lensing convergence field, $\kappa$,
can constrain non-linear galaxy bias. 

\subsection*{Phase difference distribution}

Figure \ref{histograms} shows histograms of the phase differences.
In the top panel, we show the phase difference histogram 
for the case $b_1 = 1.3$ and $c_2 = -0.2$ for three different 
bins in harmonic coefficients $l={60-79,100-119,140-149}$. The lines
display the mean value of our survey realizations, and the shaded
regions show the RMS dispersion between surveys. It can be clearly
distinguished that wave modes with lower frequency have a more 
peaked distribution than wave modes with high frequency. This makes
sense as higher wave numbers correspond to smaller physical scales,
which due to local dynamics are the first to lose coherence. 
We use information from all
the modes in our likelihood analysis. 

In the middle panel of Figure \ref{histograms} we show the 
histogram with varying values of the non-linear bias parameter 
$c_2$. All the histograms are with harmonic coefficients in the
range $l={100-119}$. It is clear that the larger the (absolute) magnitude of $c_2$ 
the broader the distribution of phase differences. This is because the 
non-linear bias smears the linear relation between matter and galaxies. 
The results are similar as well for other bins in harmonic coefficient $l$. 

In the bottom panel of Figure \ref{histograms} we show
the phase difference histograms of different combinations of bias,
but all with abs$(c_2)=0.2$, with $l={110-119}$. As in the other two panels the
lines show the mean value from our realizations of the survey,
while the filled areas show the dispersion between the different
realizations with an amplitude given by the RSM of the values. 
This panel shows how in practise the histograms depend only
on the absolute value of $c_2$, while the sign of $c_2$ and
the linear bias have almost no effect. The theoretical reasons
for this have been explained in section \ref{theory}.

\subsection*{Non-linear bias contraints}

\begin{figure*}
 \includegraphics[width=240px]{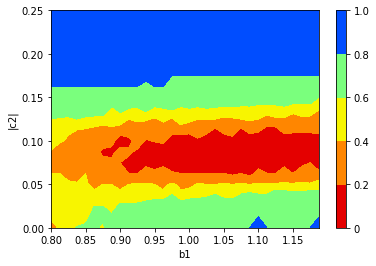} 
 \includegraphics[width=240px]{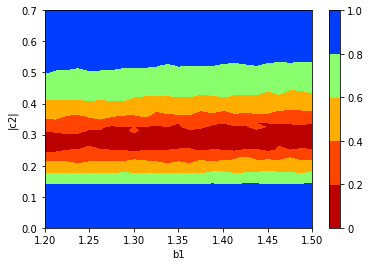}
 \includegraphics[width=240px]{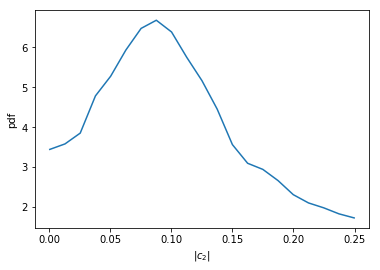} 
 \includegraphics[width=240px]{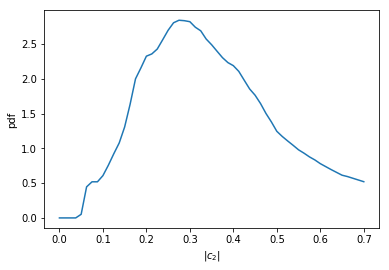}
 \caption{
  Top panels: Posterior probability contours for a simulated measurement
  of the linear ($b_1$) and non-linear ($c_2$) bias parameters
  from the phase-difference between the weak lensing convergence
  and the galaxy convergence fields on a DES-like survey of $1473$ sq. deg.
  Labeled values show the encompassed probability within contours.
  The fiducial bias values are $b_1 = 1.0$ and $c_2 = 0.1$
   (left), and $b_1=1.35$ and $c_2 =0.3$ (right). 
  It is clear that the phase
  differences constrain only $c_2$. Bottom panels: the (marginalised) probability distribution functions
  of $c_2$ for the same cases as the top panels.    
 }
 \label{likeliAA}
\end{figure*}

Figure \ref{likeliAA} shows the posterior probability contours 
for the linear and non-linear galaxy bias parameters $(b_1,c_2)$
of our simulated measurement, for two selected cases. The contour colour
values indicate the encompassed probability within the contours. 
The true bias values underlying the simulation are $b_1=1.0$ and $c_2=0.05$
for the top left panel, and $b_1=1.35$ and $c_2 = 0.3$ for the top right panel. 
Flat priors have been set on these parameters for the ranges shown in the plot. 
These plots have been generated with values on a grid with step size $\Delta b=0.0125$.  
We have checked that we get practically the same results for negative values of $c_2$.
We have combined $9$ spherical harmonic bins covering from 
$\ell=20$ to $\ell=200$.  

Clearly, as seen in these plots, there is an almost perfect degeneracy 
in the linear galaxy bias, confirming that the phase-difference
is only sensitive to the non-linear galaxy bias (see section \ref{theory}).  
From Figure \ref{likeliAA} it is also clear that the
non-linear bias $c_2$ is constrained by the phase difference
information. Marginalising over the linear bias gives 
a probability distribution function (pdf) for $|c_2|$, which
is shown in the bottom panels. The maximum likelihood values 
and standard deviation values for the two cases shown are 
$c_2=0.095$ and $\Delta c_2 = 0.064$, and $c_2 = 0.27$ and 
$\Delta c_2 = 0.145$ for the left and right panels respectively.  
From the left panel we see that it is clearly possible to 
distinguish small values of the non-linear bias with this method. 

\begin{figure}
 \includegraphics[width=250px]{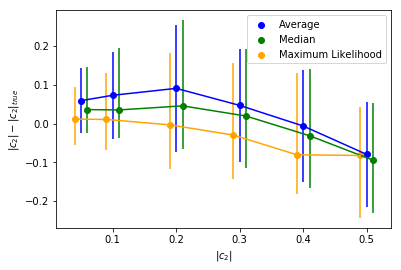}
 \caption{
  Constraining power of phase-difference histograms for $c_2$, for different 
  true values. The vertical axis shows an estimation of the systematic errors while
  the error bars show the statistical errors. We show three different 
  measurements for $c_2$ and its error: the mean and standard deviation (blue), 
  the median and the 16th to 84th percentiles region (green), and the maximum likelihood
  and the 68\% probability under equal values of the likelihood (orange).  }
 \label{constr}
\end{figure}

In Figure \ref{constr} we show the constraining power of the
phase difference histograms for different values of $c_2$. 
We have used the same flat prior space as the left panels in
Figure \ref{likeliAA}, corresponding to $b_1 = {1.2 - 1.5}$ and $c_2 = {0 - 0.7}$,
and a fiducial (observational) value of $b_1 = 1.35$, but have varied $c_2$.
While we are using a positive value of $c_2$, due to the symmetry discussed
in section $\ref{theory}$, we are actually constraining $|c_2|$. We have
checked this by examining a few cases with fiducial negative values of $c_2$, 
obtaining practically the same results. 

The plot shows the difference between the best estimation
of the non-linear bias, $\hat{c}_2$ and its true value $c_2$,
(i.e the systematic error) and the error bars correspond to the 
statistical errors on $c_2$ from the PDF marginalized over all values
of $b_1$. 
We considered three different estimations for the best fit values and errors: 
a) the mean and standard deviation, b) the median and the 16th to 84th percentile region, 
and c) the maximum likelihood and the 68\% probability range that falls under equal values of the likelihood. 
All of these give similar results and errors. 
Of the three cases, the mean value is the one that gives higher values of $c_2$;
this is due to the skewness of the PDF distribution. This case also depends on the range allowed by
the prior, increasing slightly as the prior also increases. In all three cases the systematic
errors while non-negligible are clearly smaller than the statistical errors. In turn, the
overall error allows for discriminating between different values of $c_2$. Thus
it clear that the phase differences alone have considerable constraining power.

\section{discussion and conclusion}
\label{discussion}

We have shown that the phase differences between the dark matter
convergence field and the galaxy field (weighted with a lensing kernel) 
can constrain non-linear galaxy bias. 
We can now compare the constraining power of this method with current 
results from observational data.

\textit{The Dark Energy Survey} \citep{2018PhRvD..98d3526A}: Up to now the first and only measure 
of the non-linear galaxy bias in DES has been obtained 
by fitting galaxy counts in cells from
the Science Validation Data,
consisting of 116 square degrees.
\cite{Salvador2018} looked at the moments of the counts in cells in five bins in redshift,
measuring a non-zero negative non-linear bias $b_2$. Values and errors in $b_2$ are
broad and similar at all redshifts. From their Figure 9, $c_2 = -0.75 \pm 0.55$ at $z=0.9$. We expect this error
to be considerably reduced in DES DR3; scaling the error according to the square root of the area
we are using in our simulations yields an error of $0.15$, which indicates that the counts-in-cells and phase difference methods would be competitive.

\textit{Sloan Digital Sky Survey}, SDSS-III \citep{SDSSIII}:
The non-linear galaxy bias has been measured 
from the clustering of galaxy samples
by \cite{GilMarin2017} using data from the Baryon Oscillation Spectroscopic
Survey, SDSS Data Release 12 \citep{SDSS12}. The non-linear galaxy bias $b_2$
has been obtained by fitting the bispectrum of the LOWZ (z=0.32) and CMASS (z=0.57)
galaxy samples. The authors took into account the non-local $b_s$ and the third order
non-linear $b_{nl}$ biases, which were set to their theoretically predicted values from $b_1$.
From Table 3, averaging over the two fits in the paper, and assuming no contribution
to errors from $\sigma_8$, we can deduce an approximate
value of $c_2 = 0.45 \pm 0.16$ at $z=0.3$ and $c_2 = 0.502 \pm 0.054$ at $z=0.6$.
The respective areas for these measurements are 8337 sq. deg for LOWZ and 9376 sq. deg.
for CMASS, which would yield about 2.5 larger errors when rescaled to our fiducial survey.

\textit{WiggleZ} \citep{Drinkwater2010}:
Measurements of the non-linear bias from WiggleZ
have been performed by \cite{MarinWiggleZ} by analysing the
three point clustering of bright emission galaxies in an area of 816 sq. deg,
in three redshift bins. From Table 2, we can deduce that the non-linear
bias parameter is approximately $c_2 = -0.5 \pm 0.16$ at $z=0.36$,
$c_2=-0.41 \pm 0.9$ at $z=0.55$ and $c_2=0.45 \pm 0.07$ at $z=0.66$, not
taking into account contribution to the errors from $b_1$. The statistical errors would be reduced about a factor of 34\% for an area similar to our simulations.

\textit{The Sloan Digital Sky Survey} SDSS-IV \citep{Blanton2017}: The eBOSS collaboration
have used spectroscopic data to measure the growth of structure of their luminous red galaxy sample together 
with the high redshift tail of SDSS-III CMASS galaxies, with an effective redshift of $z=0.698$ \citep{GilMarin-eboss2020},
fitting $b_2$ in the process. Their results depend on the exact galaxy sample used (north or south)
and the type of fit to the clustering data (including or not including the hexadecapole). In any case, from Table F3 and
figure E2 we can see that the equivalent values of $c_2$ range between $1.10$ and $1.47$ with large estimated errors,
which range between 87\% of the $c_2$ value and $135\%$. This result corresponds to targets covering an area 
of approximately 7500 sq. deg. (although the actual area used by the authors varies in redshift), which is about
5 times larger than our phase difference simulations. While SDSS is at a lower redshift, the larger errors seem
to indicate that our method will likely yield better results. Interestingly enough, in Figure E2 the non-linear
bias likelihood is approximately symmetrical for positive and negative values, and it is necessary to place a prior of $b_2 > 0$. 
 
We may compare the current observational results with our estimations
in Figure \ref{constr}, taking into account that they are
at different redshifts, and correcting for the different areas. 
Except for one particular redshift of WiggleZ ($z$=0.66) our 
method predicts better accuracy for our sample than all the
other methods on their galaxy samples, when scaled by the area. 
This is particularly noteworthy for SDSS-IV, where  
error in $c_2$ ranges from 1.27 to 1.69, while our 
typical errors are ten or more times smaller.

We have shown that the phase difference method is a
useful tool to measure the non-linear bias parameter 
$c_2$ independently of the linear bias $b_1$. 
Measuring the non-linear bias from the phase difference method,
together with measuring the linear bias from the large scale galaxy
clustering, will be a way to observationally falsify (or verify)  
the relationship between $c_2=b_2/b_1$ and $b_1$ obtained from simulations in the literature (eg. Figure 4, Eq 5.2 of \cite{LaBaldauf2016}, or \cite{Hoffmann2017}).
This will give insight into gravitational clustering and can act as a window to test non-standard gravitational theories. 
Note that the relationship found from simulations between $c_2$ and $b_1$ is a fit obtained in 
simulations for dark matter halos rather than galaxies themselves, so
one needs to use a model for the population of galaxies in halos, 
(for instance, a Halo Occupation Distribution) or restrict the study to 
massive halos only, where central galaxies are certain to be observed.
Another option is to assume that the relationship between $c_2$ and $b_1$
is correct for dark halos, but then use the observational measurements of the bias parameters
to gain insight into galaxy formation, i.e. how galaxies populate the halos. This could be done for galaxies split by type, luminosity and colours. 

In this paper we have produced a proof of concept 
of the utility of the phase difference statistics. 
In our model, we have made some simplifications that
may either degrade or improve our constraints. Firstly
we have fixed the cosmological parameters
and applied a local non-linear galaxy bias; when
marginalising over cosmological parameters
and non-local bias the constraints may degrade
in practice. 
In addition, we have fixed
the redshift of our sources for the lensing sample
to $z=1$. Allowing for several source redshift bins 
for a given galaxy sample of lenses will considerably improve
the constraints, as information on galaxy bias will be obtained
which would otherwise be smoothed out due the to lensing kernel. 

We now turn to the conclusions:

\begin{itemize}
\item We have shown that the phase difference between the wave modes
of the weak lensing convergence maps and the wave modes of 
the galaxy density maps integrated over the same lensing kernel
are a reliable and powerful tool to constrain non-linear galaxy bias
from observations. 
\item We have demonstrated that the non-linear bias constraints 
on the parameter $c_2 = b_2/b_1$ are independent of the constraints
on $b_1$ and that the phase differences are only sensitive to the
absolute value of $c_2$. The sign can be inferred from theory for
a given $b_1$. 
\item For a survey similar to the Dark Energy Survey, 
with measurements of convergence maps at $z=1$, and an
area of 1473 sq. deg., we found that we can constrain
$c_2$ with an error of approximately 0.1 for a wide range of $c_2$. 

\item Our phase difference method estimations of the
non-linear bias constraints are expected to provide
the strongest constraints for a given area, complementary
and competitive with current results from SDSS-III, SDSS-IV, DES and WiggleZ
observations. We expect this method will be readily applied in the
near future to DES, EUCLID and other forthcoming weak 
lensing experiments.

\end{itemize}



\section*{Acknowledgements}

MM acknowledges support from the European Union's Horizon 2020 
research and innovation program under Marie Sklodowska-Curie grant agreement No 6655919.
DB is supported by STFC grant ST/N000668/1. 
We thank Rafal Szepietowski and Enrique Gaztanaga for useful comments. 
Part of the numerical computations for this work were done on the 
Sciama High Performance Compute (HPC) cluster which is supported by 
the ICG, SEPNet and the University of Portsmouth.
This work also used the DiRAC@Durham facility managed by the Institute for Computational Cosmology on behalf of the STFC DiRAC HPC Facility (www.dirac.ac.uk). The equipment was funded by BEIS capital funding via STFC capital grants ST/K00042X/1, ST/P002293/1, ST/R002371/1 and ST/S002502/1, Durham University and STFC operations grant ST/R000832/1. DiRAC is part of the National e-Infrastructure.
Finally, we acknowldedge support from Port d'Informaci\'o Cient\'ifica (PIC), maintained through a collaboration of CIEMAT and IFAE, with additional support from Universitat Aut\`onoma de Barcelona and ERDF.




\bibliographystyle{mnras}
\bibliography{bphases} 

\begin{thebibliography}{}
\makeatletter
\relax
\def\mn@urlcharsother{\let\do\@makeother \do\$\do\&\do\#\do\^\do\_\do\%\do\~}
\def\mn@doi{\begingroup\mn@urlcharsother \@ifnextchar [ {\mn@doi@}
  {\mn@doi@[]}}
\def\mn@doi@[#1]#2{\def\@tempa{#1}\ifx\@tempa\@empty \href
  {http://dx.doi.org/#2} {doi:#2}\else \href {http://dx.doi.org/#2} {#1}\fi
  \endgroup}
\def\mn@eprint#1#2{\mn@eprint@#1:#2::\@nil}
\def\mn@eprint@arXiv#1{\href {http://arxiv.org/abs/#1} {{\tt arXiv:#1}}}
\def\mn@eprint@dblp#1{\href {http://dblp.uni-trier.de/rec/bibtex/#1.xml}
  {dblp:#1}}
\def\mn@eprint@#1:#2:#3:#4\@nil{\def\@tempa {#1}\def\@tempb {#2}\def\@tempc
  {#3}\ifx \@tempc \@empty \let \@tempc \@tempb \let \@tempb \@tempa \fi \ifx
  \@tempb \@empty \def\@tempb {arXiv}\fi \@ifundefined
  {mn@eprint@\@tempb}{\@tempb:\@tempc}{\expandafter \expandafter \csname
  mn@eprint@\@tempb\endcsname \expandafter{\@tempc}}}

\bibitem[\protect\citeauthoryear{{Abbott} et~al.,}{{Abbott}
  et~al.}{2018}]{2018PhRvD..98d3526A}
{Abbott} T.~M.~C.,  et~al., 2018, \mn@doi [\prd] {10.1103/PhysRevD.98.043526},
  \href {https://ui.adsabs.harvard.edu/abs/2018PhRvD..98d3526A} {98, 043526}

\bibitem[\protect\citeauthoryear{{Alam} et~al.,}{{Alam} et~al.}{2015}]{SDSS12}
{Alam} S.,  et~al., 2015, \mn@doi [\apjs] {10.1088/0067-0049/219/1/12}, \href
  {https://ui.adsabs.harvard.edu/abs/2015ApJS..219...12A} {219, 12}

\bibitem[\protect\citeauthoryear{{Ali}, {Obreschkow}, {Howlett}, {Bonvin},
  {Llinares}, {Oliveira Franco}  \& {Power}}{{Ali} et~al.}{2018}]{Kamran2018}
{Ali} K.,  {Obreschkow} D.,  {Howlett} C.,  {Bonvin} C.,  {Llinares} C.,
  {Oliveira Franco} F.,   {Power} C.,  2018, \mn@doi [\mnras]
  {10.1093/mnras/sty1696}, 479, 2743

\bibitem[\protect\citeauthoryear{{Baldauf}, {Seljak}, {Desjacques}  \&
  {McDonald}}{{Baldauf} et~al.}{2012}]{Baldauf12}
{Baldauf} T.,  {Seljak} U.,  {Desjacques} V.,   {McDonald} P.,  2012, \mn@doi
  [\prd] {10.1103/PhysRevD.86.083540}, \href
  {http://adsabs.harvard.edu/abs/2012PhRvD..86h3540B} {86, 083540}

\bibitem[\protect\citeauthoryear{{Bartelmann} \& {Schneider}}{{Bartelmann} \&
  {Schneider}}{2001}]{2001PhR...340..291B}
{Bartelmann} M.,  {Schneider} P.,  2001, \mn@doi [\physrep]
  {10.1016/S0370-1573(00)00082-X}, \href
  {https://ui.adsabs.harvard.edu/abs/2001PhR...340..291B} {340, 291}

\bibitem[\protect\citeauthoryear{{Bel}, {Hoffmann}  \& {Gazta{\~n}aga}}{{Bel}
  et~al.}{2015}]{BelHoffmannGazta15}
{Bel} J.,  {Hoffmann} K.,   {Gazta{\~n}aga} E.,  2015, \mn@doi [\mnras]
  {10.1093/mnras/stv1600}, \href
  {http://adsabs.harvard.edu/abs/2015MNRAS.453..259B} {453, 259}

\bibitem[\protect\citeauthoryear{{Blanton} et~al.,}{{Blanton}
  et~al.}{2017}]{Blanton2017}
{Blanton} M.~R.,  et~al., 2017, \mn@doi [\aj] {10.3847/1538-3881/aa7567}, \href
  {https://ui.adsabs.harvard.edu/abs/2017AJ....154...28B} {154, 28}

\bibitem[\protect\citeauthoryear{{Byun}, {Eggemeier}, {Regan}, {Seery}  \&
  {Smith}}{{Byun} et~al.}{2017}]{ByunLCF}
{Byun} J.,  {Eggemeier} A.,  {Regan} D.,  {Seery} D.,   {Smith} R.~E.,  2017,
  \mn@doi [\mnras] {10.1093/mnras/stx1681}, \href
  {https://ui.adsabs.harvard.edu/abs/2017MNRAS.471.1581B} {471, 1581}

\bibitem[\protect\citeauthoryear{{Chan}, {Scoccimarro}  \& {Sheth}}{{Chan}
  et~al.}{2012}]{Chan12}
{Chan} K.~C.,  {Scoccimarro} R.,   {Sheth} R.~K.,  2012, \mn@doi [\prd]
  {10.1103/PhysRevD.85.083509}, \href
  {http://adsabs.harvard.edu/abs/2012PhRvD..85h3509C} {85, 083509}

\bibitem[\protect\citeauthoryear{{Chiang} \& {Coles}}{{Chiang} \&
  {Coles}}{2000}]{ChiangColes2000}
{Chiang} L.-Y.,  {Coles} P.,  2000, \mn@doi [\mnras]
  {10.1046/j.1365-8711.2000.03086.x}, \href
  {http://adsabs.harvard.edu/abs/2000MNRAS.311..809C} {311, 809}

\bibitem[\protect\citeauthoryear{{Chiang}, {Coles}  \& {Naselsky}}{{Chiang}
  et~al.}{2002}]{ChiangColes2002}
{Chiang} L.-Y.,  {Coles} P.,   {Naselsky} P.,  2002, \mn@doi [\mnras]
  {10.1046/j.1365-8711.2002.05931.x}, \href
  {http://adsabs.harvard.edu/abs/2002MNRAS.337..488C} {337, 488}

\bibitem[\protect\citeauthoryear{{Chiang}, {Naselsky}  \& {Coles}}{{Chiang}
  et~al.}{2004}]{ChiangNaselsky2004}
{Chiang} L.-Y.,  {Naselsky} P.~D.,   {Coles} P.,  2004, \mn@doi [\apjl]
  {10.1086/382211}, \href {http://adsabs.harvard.edu/abs/2004ApJ...602L...1C}
  {602, L1}

\bibitem[\protect\citeauthoryear{{Desjacques}, {Jeong}  \&
  {Schmidt}}{{Desjacques} et~al.}{2018}]{2018PhR...733....1D}
{Desjacques} V.,  {Jeong} D.,   {Schmidt} F.,  2018, \mn@doi [\physrep]
  {10.1016/j.physrep.2017.12.002}, \href
  {https://ui.adsabs.harvard.edu/abs/2018PhR...733....1D} {733, 1}

\bibitem[\protect\citeauthoryear{{Drinkwater} et~al.,}{{Drinkwater}
  et~al.}{2010}]{Drinkwater2010}
{Drinkwater} M.~J.,  et~al., 2010, \mn@doi [\mnras]
  {10.1111/j.1365-2966.2009.15754.x}, \href
  {https://ui.adsabs.harvard.edu/abs/2010MNRAS.401.1429D} {401, 1429}

\bibitem[\protect\citeauthoryear{{Efron} \& {Tibshirani}}{{Efron} \&
  {Tibshirani}}{1993}]{bootstrapbook}
{Efron} B.,  {Tibshirani} R.~J.,  1993, {An Introduction to the Bootstrap}.
Chapman and Hall/CRC

\bibitem[\protect\citeauthoryear{{Eggemeier} \& {Smith}}{{Eggemeier} \&
  {Smith}}{2017}]{Eggemeier2017}
{Eggemeier} A.,  {Smith} R.~E.,  2017, \mn@doi [\mnras]
  {10.1093/mnras/stw3249}, \href
  {https://ui.adsabs.harvard.edu/abs/2017MNRAS.466.2496E} {466, 2496}

\bibitem[\protect\citeauthoryear{{Eisenstein} et~al.,}{{Eisenstein}
  et~al.}{2011}]{SDSSIII}
{Eisenstein} D.~J.,  et~al., 2011, \mn@doi [\aj] {10.1088/0004-6256/142/3/72},
  \href {https://ui.adsabs.harvard.edu/abs/2011AJ....142...72E} {142, 72}

\bibitem[\protect\citeauthoryear{{Gaztanaga} \& {Frieman}}{{Gaztanaga} \&
  {Frieman}}{1994}]{1994ApJ...437L..13G}
{Gaztanaga} E.,  {Frieman} J.~A.,  1994, \mn@doi [\apjl] {10.1086/187671},
  \href {https://ui.adsabs.harvard.edu/abs/1994ApJ...437L..13G} {437, L13}

\bibitem[\protect\citeauthoryear{{Gil-Mar{\'\i}n}, {Nore{\~n}a}, {Verde},
  {Percival}, {Wagner}, {Manera}  \& {Schneider}}{{Gil-Mar{\'\i}n}
  et~al.}{2015}]{2015MNRAS.451..539G}
{Gil-Mar{\'\i}n} H.,  {Nore{\~n}a} J.,  {Verde} L.,  {Percival} W.~J.,
  {Wagner} C.,  {Manera} M.,   {Schneider} D.~P.,  2015, \mn@doi [\mnras]
  {10.1093/mnras/stv961}, \href
  {https://ui.adsabs.harvard.edu/abs/2015MNRAS.451..539G} {451, 539}

\bibitem[\protect\citeauthoryear{{Gil-Mar{\'{\i}}n}, {Percival}, {Verde},
  {Brownstein}, {Chuang}, {Kitaura}, {Rodr{\'{\i}}guez-Torres}  \&
  {Olmstead}}{{Gil-Mar{\'{\i}}n} et~al.}{2017}]{GilMarin2017}
{Gil-Mar{\'{\i}}n} H.,  {Percival} W.~J.,  {Verde} L.,  {Brownstein} J.~R.,
  {Chuang} C.-H.,  {Kitaura} F.-S.,  {Rodr{\'{\i}}guez-Torres} S.~A.,
  {Olmstead} M.~D.,  2017, \mn@doi [\mnras] {10.1093/mnras/stw2679}, \href
  {http://adsabs.harvard.edu/abs/2017MNRAS.465.1757G} {465, 1757}

\bibitem[\protect\citeauthoryear{{Gil-Mar{\'\i}n} et~al.,}{{Gil-Mar{\'\i}n}
  et~al.}{2020}]{GilMarin-eboss2020}
{Gil-Mar{\'\i}n} H.,  et~al., 2020, arXiv e-prints, \href
  {https://ui.adsabs.harvard.edu/abs/2020arXiv200708994G} {p. arXiv:2007.08994}

\bibitem[\protect\citeauthoryear{{G{\'o}rski}, {Hivon}, {Banday}, {Wand elt},
  {Hansen}, {Reinecke}  \& {Bartelmann}}{{G{\'o}rski}
  et~al.}{2005}]{2005ApJ...622..759G}
{G{\'o}rski} K.~M.,  {Hivon} E.,  {Banday} A.~J.,  {Wand elt} B.~D.,  {Hansen}
  F.~K.,  {Reinecke} M.,   {Bartelmann} M.,  2005, \mn@doi [\apj]
  {10.1086/427976}, \href
  {https://ui.adsabs.harvard.edu/abs/2005ApJ...622..759G} {622, 759}

\bibitem[\protect\citeauthoryear{{Hoffmann}, {Bel}  \&
  {Gazta{\~n}aga}}{{Hoffmann} et~al.}{2017}]{Hoffmann2017}
{Hoffmann} K.,  {Bel} J.,   {Gazta{\~n}aga} E.,  2017, \mn@doi [\mnras]
  {10.1093/mnras/stw2876}, \href
  {https://ui.adsabs.harvard.edu/abs/2017MNRAS.465.2225H} {465, 2225}

\bibitem[\protect\citeauthoryear{{Howlett}, {Manera}  \& {Percival}}{{Howlett}
  et~al.}{2015}]{lpicola15}
{Howlett} C.,  {Manera} M.,   {Percival} W.~J.,  2015, \mn@doi [Astronomy and
  Computing] {10.1016/j.ascom.2015.07.003}, \href
  {http://adsabs.harvard.edu/abs/2015A%26C....12..109H} {12, 109}

\bibitem[\protect\citeauthoryear{{Lazeyras}, {Wagner}, {Baldauf}  \&
  {Schmidt}}{{Lazeyras} et~al.}{2016}]{LaBaldauf2016}
{Lazeyras} T.,  {Wagner} C.,  {Baldauf} T.,   {Schmidt} F.,  2016, \mn@doi
  [\jcap] {10.1088/1475-7516/2016/02/018}, \href
  {https://ui.adsabs.harvard.edu/abs/2016JCAP...02..018L} {2016, 018}

\bibitem[\protect\citeauthoryear{{Manera} et~al.,}{{Manera}
  et~al.}{2013}]{pthalos13}
{Manera} M.,  et~al., 2013, \mn@doi [\mnras] {10.1093/mnras/sts084}, \href
  {http://adsabs.harvard.edu/abs/2013MNRAS.428.1036M} {428, 1036}

\bibitem[\protect\citeauthoryear{{Manera} et~al.,}{{Manera}
  et~al.}{2015}]{pthalos15}
{Manera} M.,  et~al., 2015, \mn@doi [\mnras] {10.1093/mnras/stu2465}, \href
  {http://adsabs.harvard.edu/abs/2015MNRAS.447..437M} {447, 437}

\bibitem[\protect\citeauthoryear{{Mar{\'\i}n} et~al.,}{{Mar{\'\i}n}
  et~al.}{2013}]{MarinWiggleZ}
{Mar{\'\i}n} F.~A.,  et~al., 2013, \mn@doi [\mnras] {10.1093/mnras/stt520},
  \href {https://ui.adsabs.harvard.edu/abs/2013MNRAS.432.2654M} {432, 2654}

\bibitem[\protect\citeauthoryear{{Martin}}{{Martin}}{1990}]{jackknifeFAILS}
{Martin} M.~M.,  1990, \mn@doi [Canadian Journal of Statistics]
  {10.2307/3315563}, 18, 149

\bibitem[\protect\citeauthoryear{{Prat} et~al.,}{{Prat}
  et~al.}{2018}]{2018PhRvD..98d2005P}
{Prat} J.,  et~al., 2018, \mn@doi [\prd] {10.1103/PhysRevD.98.042005}, \href
  {https://ui.adsabs.harvard.edu/abs/2018PhRvD..98d2005P} {98, 042005}

\bibitem[\protect\citeauthoryear{{Salvador} et~al.,}{{Salvador}
  et~al.}{2019}]{Salvador2018}
{Salvador} A.~I.,  et~al., 2019, \mn@doi [\mnras] {10.1093/mnras/sty2802},
  \href {https://ui.adsabs.harvard.edu/abs/2019MNRAS.482.1435S} {482, 1435}

\bibitem[\protect\citeauthoryear{{Szepietowski}, {Bacon}, {Dietrich}, {Busha},
  {Wechsler}  \& {Melchior}}{{Szepietowski}
  et~al.}{2014}]{SzepietowskiBacon2014}
{Szepietowski} R.~M.,  {Bacon} D.~J.,  {Dietrich} J.~P.,  {Busha} M.,
  {Wechsler} R.,   {Melchior} P.,  2014, \mn@doi [\mnras]
  {10.1093/mnras/stu380}, \href
  {http://adsabs.harvard.edu/abs/2014MNRAS.440.2191S} {440, 2191}

\bibitem[\protect\citeauthoryear{{Tassev}, {Zaldarriaga}  \&
  {Eisenstein}}{{Tassev} et~al.}{2013}]{cola13}
{Tassev} S.,  {Zaldarriaga} M.,   {Eisenstein} D.~J.,  2013, \mn@doi [\jcap]
  {10.1088/1475-7516/2013/06/036}, \href
  {http://adsabs.harvard.edu/abs/2013JCAP...06..036T} {6, 036}

\bibitem[\protect\citeauthoryear{{Verde} et~al.,}{{Verde}
  et~al.}{2002}]{2002MNRAS.335..432V}
{Verde} L.,  et~al., 2002, \mn@doi [\mnras] {10.1046/j.1365-8711.2002.05620.x},
  \href {https://ui.adsabs.harvard.edu/abs/2002MNRAS.335..432V} {335, 432}

\bibitem[\protect\citeauthoryear{{Watts}, {Coles}  \& {Melott}}{{Watts}
  et~al.}{2003}]{WattsColes2003}
{Watts} P.,  {Coles} P.,   {Melott} A.,  2003, \mn@doi [\apjl]
  {10.1086/376351}, \href {http://adsabs.harvard.edu/abs/2003ApJ...589L..61W}
  {589, L61}

\bibitem[\protect\citeauthoryear{{Wolstenhulme}, {Bonvin}  \&
  {Obreschkow}}{{Wolstenhulme} et~al.}{2015}]{WolstenhulmeBoyin2015}
{Wolstenhulme} R.,  {Bonvin} C.,   {Obreschkow} D.,  2015, \mn@doi [\apj]
  {10.1088/0004-637X/804/2/132}, \href
  {http://adsabs.harvard.edu/abs/2015ApJ...804..132W} {804, 132}

\bibitem[\protect\citeauthoryear{{Zuntz} et~al.,}{{Zuntz}
  et~al.}{2018}]{2018MNRAS.481.1149Z}
{Zuntz} J.,  et~al., 2018, \mn@doi [\mnras] {10.1093/mnras/sty2219}, \href
  {https://ui.adsabs.harvard.edu/abs/2018MNRAS.481.1149Z} {481, 1149}

\bibitem[\protect\citeauthoryear{{de la Torre} \& {Peacock}}{{de la Torre} \&
  {Peacock}}{2013}]{delaTorre2013}
{de la Torre} S.,  {Peacock} J.~A.,  2013, \mn@doi [\mnras]
  {10.1093/mnras/stt1333}, \href
  {http://adsabs.harvard.edu/abs/2013MNRAS.435..743D} {435, 743}

\makeatother
\end{thebibliography}




\appendix

\section{$c_2$ symmetriy}
\label{APc2}

At leading order in the spherical harmonic coefficients, there is a symmetry
in the phase differences between the galaxy and the weak lensing convergence 
fields, that entails that we cannot distinguish between a positive and a negative
value of $c_2$, the second order perturbation theory coefficient. In this 
section we demonstrate this symmetry in practice. 

\begin{figure}
 \includegraphics[width=200px]{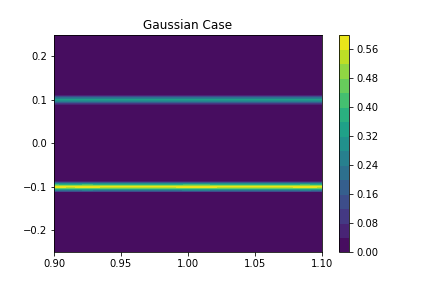}
 \caption{Likelihood for Gaussian realizations with $c_2=-0.1$.}
 \label{gaussiansymmetry}
\end{figure}

\begin{figure}
 \includegraphics[width=200px]{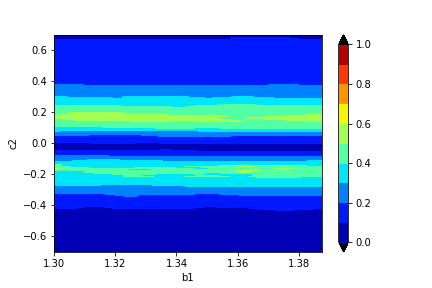}
 \caption{Likelihood for a DES-like Y1 survey with $c2 = -0.15$ from simulations. 
Because of the symmetry in $c_2$ we can only constrain its absolute value.}
 \label{c2symmetrystandard}
\end{figure}

Figure \ref{gaussiansymmetry} shows the likelihood results for a simple test case in which we obtain 
a full sky healpix map; for each pixel we draw a value from
a normal distribution $g \sim \mathcal{N}(0,1)$, then calculate a non-linear value $g' = b_1 (g + 0.5 c_2 g^2)$.
For each combination of $b_1$ and $c_2$ values we obtain the phase difference histogram between the 
linear and non-linear maps (representing the lensing and galaxy convergence maps respectively). Choosing as our `observation' a fiducial value of $c_2 = -0.1$, and running 50 realizations
of this case to use for our statistics, we obtain the likelihood values (via the process described in section \ref{statistics}) for each of the parameter combinations.
The results show that we cannot distinguish between $c_2 = 0.1$ and the true $c_2 = -0.1$ value, as expected from equation \ref{eq:symmetry}. 
We have used healpix $nside=512$ and the same $\ell$ bins as in the main paper. 
 
Figure \ref{c2symmetrystandard} shows the likelihood for results with fiducial $c2 = -0.15$ and a DES-like Y1 survey (see section \ref{sims} for our prescription). We allow for positive and negative values of $c_2$; again we clearly see a symmetry between positive and negative of the non-linear contribution.  For the statistics used, see the explanation in Section \ref{statistics}.

\section{Minimum redshift from discretisation}

\label{kg-appendix}

The fact that galaxies are discrete objects impacts the estimation of the
galaxy convergence field, $\kappa_g$. At low redshifts, because of the very few number of galaxies,
healpix pixels may be assigned such large overdensity values that the
resulting galaxy convergence computed from them stops being a meaningful tracer of the dark matter field.
In order to mitigate these discretization effects we set a minimum redshift of $z=0.05$
from which to compute our galaxy convergence field. 

\begin{figure}
 \includegraphics[width=250px]{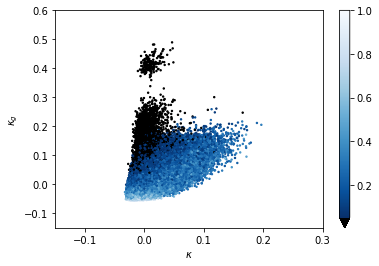}
\caption{Scatter plot between $\kappa_g$ and $\kappa$ for sources at redshift 1.0.  Colour values represent the lowest redshift of a galaxy in each pixel. }
\label{scatter_zmin}
\end{figure}

Figure \ref{scatter_zmin} shows a scatter plot between the dark matter convergence, $\kappa$,
and the galaxy convergence, $\kappa_g$, where galaxies have been assigned by randomly subsampling
dark matter particles. We used approximately 800 sq. deg. from our PICOLA simulation, using
4M healpix pixels as lines of sight. 
The number density of galaxies subsampling the dark matter has been chosen to be similar to
the number of galaxies of DES up to $z=1$. The colors in the scatter plot show the 
minimum redshift of galaxies in the line of sight pixels for which we computed 
the $\kappa_g(z=1)$. Black points are those pixels where the closest galaxy 
has a minumum redshift below $0.05$. These pixels have $\kappa_g$ values well off the expected linear 
relationship between $\kappa$ and $\kappa_g$ for galaxies subsampling the dark matter field. 
These misleading values come from the discretization effects described above; 
we therefore set a minimum redshift of $z=0.05$ from which to compute $\kappa_g$. 
The effect of this cut on the true convergence signal is very small, as the weak lensing kernel 
peaks at $z\sim 0.5$ for sources at $z=1$ and goes to zero at the observer and source redshifts.


\bsp	
\label{lastpage}
\end{document}